\documentclass{article}
\usepackage{spconf,amsmath,graphicx}
\usepackage{multirow}
\usepackage{booktabs}
\usepackage{subfig}
\usepackage{graphicx}
\usepackage{epstopdf}
\usepackage{epsfig}
\usepackage{amsmath} 

\title{VE-KWS: Visual Modality Enhanced End-to-End Keyword Spotting}
\name{
\begin{tabular}{c}
\it Ao Zhang$^1$, He Wang$^1$, Pengcheng Guo$^1$, Yihui Fu$^1$, Lei Xie$^{1*}$\thanks{* Corresponding author. This paper was supported by MoE-CMCC ``Artificial Intelligence" Project (MCM20190701).},\\
\it Yingying Gao$^2$,  Shilei Zhang$^2$,  Junlan Feng$^2$
\end{tabular}
}
\address{
    $^1$Audio, Speech and Language Processing Group (ASLP@NPU), School of Computer Science,\\ Northwestern Polytechnical University, Xi’an, China \\
    $^2$China Mobile Research Institute, Beijing, China}
\begin{document}
\ninept
\maketitle
\begin{abstract}
\vspace{-2pt}

The performance of the keyword spotting (KWS) system based on audio modality, commonly measured in false alarms and false rejects, degrades significantly under the far field and noisy conditions. 
Therefore, audio-visual keyword spotting, which leverages complementary relationships over multiple modalities, has recently gained much attention.
However, current studies mainly focus on combining the exclusively learned representations of different modalities, instead of exploring the modal relationships during each respective modeling.
In this paper, we propose a novel visual modality enhanced end-to-end KWS framework (VE-KWS), which fuses audio and visual modalities from two aspects.
The first one is utilizing the speaker location information obtained from the lip region in videos to assist the training of multi-channel audio beamformer. 
By involving the beamformer as an audio enhancement module, the acoustic distortions, caused by the far field or noisy environments, could be significantly suppressed.
The other one is conducting cross-attention between different modalities to capture the inter-modal relationships and help the representation learning of each modality.
Experiments on the MSIP challenge corpus show that our proposed model achieves a 2.79\% false rejection rate and a 2.95\% false alarm rate on the Eval set, resulting in a new SOTA performance compared with the top-ranking systems in the ICASSP2022 MISP challenge.
\end{abstract}
\vspace{-6pt}
\section{Introduction}
\label{sec:intro}
\vspace{-3pt}
With the proliferation of mobile and intelligent devices, such as smart speakers, voice-enabled user interfaces play an increasingly crucial role in achieving natural user experiences. To achieve a hands-free speech recognition experience, keyword spotting (KWS) is considered to be one of the frontier components, which takes the responsibility to trigger the voice assistants to initiate speech recognition~\cite{kws_0}. Therefore, the prediction accuracy of KWS strongly impacts the user experience of voice assistants. 
Recent works on KWS have gained tremendous success and the KWS systems based on audio modality usually perform well under clean-speech
conditions~\cite{kwsrelate0,kwsrelate1,kwsrelate2}. 
However, their performance may degrade significantly under noisy conditions due to the interference in signal transmission and the complexity of the acoustic environment ~\cite{hard1,hard2,hard3}. To improve the fluency of human-computer interaction, a robust KWS system is important.

In recent years, many studies have been done on improving the noise robustness of KWS systems, including introducing speech enhancement modules and investigating novel network architectures. 
Researchers also utilize many training techniques, like data augmentation and semi-supervised learning, to mitigate performance deterioration under noisy speech conditions. 
The multi-look KWS proposed in~\cite{hard3} introduced a multi-look enhancement network (MLENet) to improve KWS performance in noisy and far-field conditions. 
In~\cite{convmixer}, a novel convolutional networks (CNN) encoder was proposed to make networks focus on useful spatial information, and the curriculum multi-condition training based on signal-to-noise ratio (SNR) level was applied to achieve better noise robustness. 
\begin{figure*}[t]
    \centering
    \includegraphics[width=0.8\linewidth]{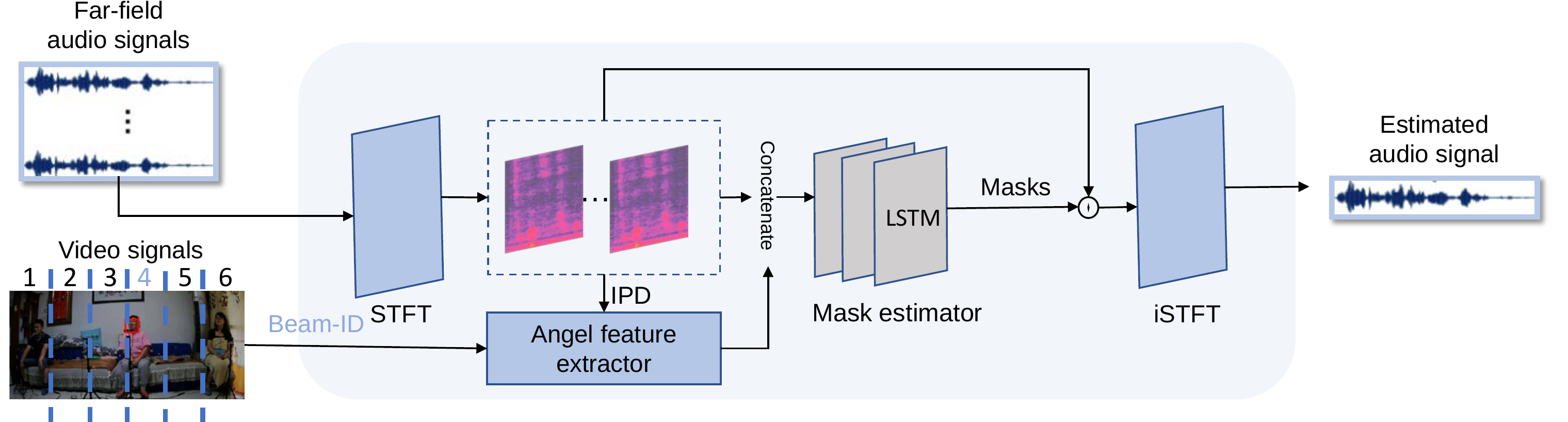}
    \vspace{-10pt}
    \caption{Block diagram of the proposed visual-assisted MVDR.}
    \label{fig:fe}
    \vspace{-15pt}
\end{figure*}

Despite the above research progress, KWS in realistic environments with a low SNR is still challenging. Since signals from audio modality can be susceptible to contamination by the complex environment, multi-modal based methods have been well explored in various tasks, including emotion recognition~\cite{emotion}, scene classification~\cite{scene}, etc., aiming to provide complementary information for the systems. These studies have proven that the introduction of visual modality can improve the robustness of systems. Inspired by this, visual modality was first introduced to the KWS task in~\cite{avkws}, which detected keywords from both audio and visual streams by computing the similarity map between features from different modalities and fed the multi-modal features into the network to make the final prediction. 
Although the audio-visual method is a promising solution for robust KWS in noisy environments, the lack of public audio-visual KWS datasets has hindered the development of this field. 
In the Multimodal Information Based Speech Processing (MISP) Challenge 2021~\cite{misp}, a sizable wake word spotting database was publicly released, offering the community a standard testbed for evaluating the audio-visual KWS systems.
The dataset consists of distant multi-microphone conversational audio-vision data collected in real-world home TV scenarios. In the challenge, the real domestic background noises, like TV or air conditions, make it hard to estimate speakers’ location and detect keywords only based on audio modality, while the visual occlusion limits the system development on visual modality too.
Thus, many participants explored the usage of multi-modal information~\cite{dku,weilai}. In~\cite{dku}, Cheng et al. proposed a fusion method called hierarchical modality aggregation (HMA), which combined intermediate feature maps of uni-modal encoders at different levels to make feature-level fusion. The method proposed in~\cite{weilai} explored a feature-level fusion by concatenating the output of uni-modal encoders and a decision-level fusion by averaging scores of systems based on audio, visual, and audio-visual.

However, most of the above methods focus on learning multi-modal information by simply concatenating different modal features from their exclusive networks. In addition, in the process of multi-channel speech enhancement, the use of visual modality is not explored. 
In this paper, we proposed a novel visual modality enhanced end-to-end KWS framework (VE-KWS), which utilizes visual modality to assist speech enhancement and adopts a cross-attention based method to learn cross-modal features. 
Specifically, for speech enhancement, we use the coordinate of the lip region in the video to compute the angle feature and guide the multi-channel speech beamformer training.
For cross-modal feature learning, firstly, a cross-attention module is introduced in the Transformer layer to capture the complementary information from other modalities, i.e. visual-aware representation for audio modality, for enhancing the learned uni-modal features.
After that, we use a bilinear layer and learnable matrixes to fuse the enhanced uni-modal features of different modalities.
Experiments on the MISP challenge corpus show that our proposed audio-visual KWS system achieves a 2.79\% false rejection rate (FRR) and a 2.95\% false alarm rate (FAR) on the Eval set, which is superior to the best system in the challenge, resulting in a new SOTA performance.
\vspace{-10pt}


\section{Proposed Methods}
\label{sec:method}


\vspace{-6pt}
\subsection{Visual-Assisted MVDR}
\vspace{-6pt}
The direction information of the speaker is important to enhance the audio signal, which has been proven in many studies~\cite{doi1,doi2,doi3}. Nevertheless, the knowledge of the actual target speaker's direction is not available in real applications. It is difficult to accurately estimate the target speaker's direction of arrival (DOA). 
However, in the multi-modal scenario, we can get the lip region from the video, which indicates the target speaker's location.
Therefore, we incorporate the lip region location information into the training of our minimum variance distortionless response (MVDR) beamformer~\cite{mvdr}.
The overall flowchart of the visual-assisted MVDR is shown in Fig.~\ref{fig:fe}. Our front-end processing module consists of an angle feature extractor, a multi-channel speech enhancement network, and a mask-based minimum variance distortionless response (MVDR) beamformer.

We first equally split each video frame into several regions along the horizontal axis and index them from left to right.
The region id of each speaker could be obtained according to the coordinate of the lip region of interest (ROI). 
Then, we compute the steering vector $\Delta_\theta$ of the corresponding region's central angle $\theta$ according to the topology of the microphone array. 
Besides, the inter-channel phase difference (IPD) is also calculated according to the phase of the observed signal of the certificate microphone pair. Finally, let $i$ and $j$ denote the microphone index and ${y}_{t, f}$ denotes the corresponding observed multi-channel signal on the frequency domain, the angle feature (AF) can be formulated as
\vspace{-5pt}
\begin{align}
\text{IPD}_{ij} &=\angle \mathbf{y}_{t, f}^i-\angle \mathbf{y}_{t, f}^j  \\
\text{AF}_{ij, \theta} &=\cos \left(\text{IPD}_{i j}-\Delta_\theta\right)
\end{align}

Secondly, we concatenate the magnitude of the first channel of the observed signal and the angle feature calculated by certificate microphone pairs along the frequency axis as the input of the speech enhancement network. For the mask estimator, we use the ideal ratio mask (IRM) of both speech and noise signals as the training target and mean square error (MSE) as the loss function.

Finally, in the MVDR step, we calculate the covariance matrices of speech and noise signals respectively based on masks estimated in the second stage via
\vspace{-10pt}
\begin{align}
\label{lab:mvdr1}
\mathbf{R}_f^{S}&=\frac{1}{\sum_t m_{t, f}^{S}} \sum_t m_{t, f}^{S} \mathbf{y}_{t, f} \mathbf{y}_{t, f}^H, \\
\mathbf{R}_f^{N}&=\frac{1}{\sum_t m_{t, f}^{N}} \sum_t m_{t, f}^{N} \mathbf{y}_{t, f} \mathbf{y}_{t, f}^H,
\end{align}
where $m_{t, f}^{S}$ and $m_{t, f}^{N}$ refer to speech and noise mask. The filter coefficients of the MVDR for speech $w_{speech}$ are derived as follows

\begin{equation}
\mathbf{w}_f=\frac{\left(\mathbf{R}_f^{N}\right)^{-1} \mathbf{R}_f^{S}}{\operatorname{tr}\left(\left(\mathbf{R}_f^{N}\right)^{-1} \mathbf{R}_f^{S}\right)} \mathbf{u}_f,
\end{equation}
where tr(·) is the matrix trace operation and $u_f$ is a one-hot vector indicating the reference microphone. Thus, we can get the final enhanced result for speech via

\begin{equation}
    \hat{\mathbf{y}}_{t, f}=\mathbf{w}_f^{\mathrm{H}} \mathbf{y}_{t, f} .
\end{equation}

We use the enhanced result as the input for following keyword spotting  to improve the robustness of the system.
\vspace{-10pt}

\subsection{Cross-Modal Feature Learning}
\vspace{-2pt}
\subsubsection{Uni-Modal Feature Extraction}
\vspace{-2pt}
The speaker's lip movements in videos carry information pertinent to both appearance and temporal dynamics. It's important to efficiently model these spatial and temporal cues when building a robust KWS system. ResNet3D proposed in~\cite{resnet3d} is proven to be effective in capturing the spatio-temporal dynamics in videos and has been widely used in different tasks, like video action recognition~\cite{resnet3dar}, speaker identification~\cite{speaker}, etc. Therefore, we build our feature extractor based on ResNet3D and adopt the same architecture for different modalities in order to keep the consistency of feature dimensions.

For visual modality, the lip regions from video frames are regarded as the input, while for audio modality, the log-Mel Filterbank (FBank) feature of each frame is spliced with adjacent context frames to form a 2-D input feature.

\vspace{-10pt}
\subsubsection{Attention-based Cross-Modal Feature Fusion}\label{Sec:2-2-2}
\begin{figure}[t]
    \centering
    \includegraphics[width=0.65\linewidth]{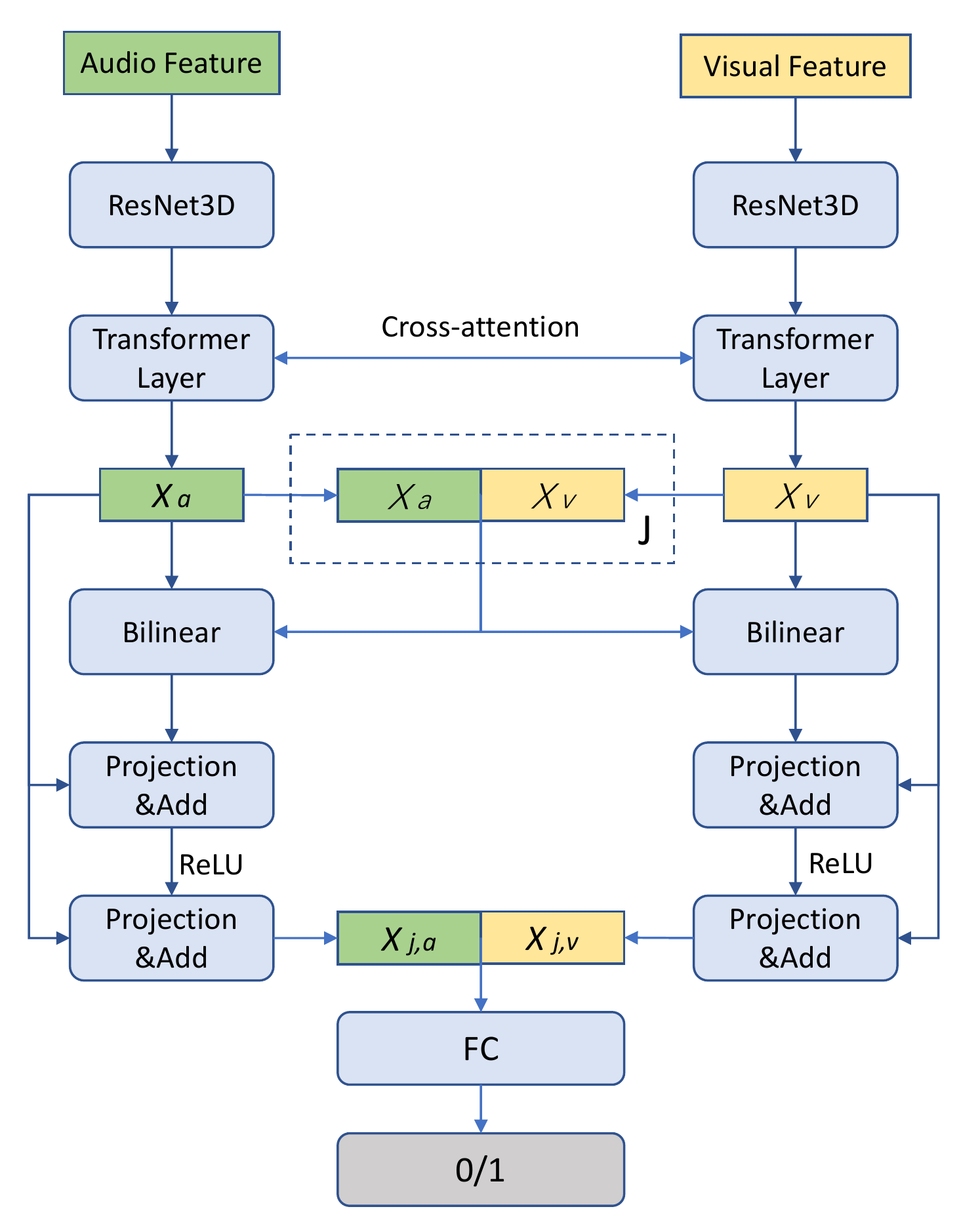}
    \caption{Block diagram of the proposed audio-visual fusion model.}
    \vspace{-5pt}
    \label{fig:fusion}
    \vspace{-10pt}
\end{figure}

\vspace{-2pt}
For the task of KWS, audio modality is more reliable than visual modality under a silent environment since the audio-based systems usually outperform their visual-based counterparts.
But the system based on audio may perform badly in a noisy environment, while visual modality is not affected by acoustic noise. Therefore, it is essential to leverage diverse and complementary relationships over multiple modalities in order to improve the robustness of the KWS system. However, the features extracted from different modalities have diverse characteristics, which makes the simple concatenation strategy the best choice. In order to reliably fuse these modalities to KWS, we first utilize a cross-attention mechanism to enhance the representation learning of each modality by incorporating information from other modalities. Then, the enhanced representations are fused by a bilinear layer and multiple residual projection layers. A block diagram of the proposed method is shown in \ref{fig:fusion}.

We employ a Transformer layer, which is efficient in extracting context-dependent representations~\cite{attisneed}, on the output of ResNet3D. Our Transformer layer consists of a self-attention module and a cross-attention module. The self-attention is employed to model the intra-modal dependencies, while the cross-attention regards the representation of other modalities as the query vector to capture the inter-model relationship. After that, we adopt a fusion method similar to~\cite{emotion} to further fuse two modalities.

Let $\mathbf{X}_a$ and $\mathbf{X}_v$ represent the enhanced audio and visual representations extracted from their Transformer layers, we first concatenate them on the last dim to form the joint representation $\mathbf{J}$. After that, we use bilinear pooling, a powerful method for fusing multi-modal representations~\cite{bilinear}, to fuse the joint representation and uni-modal representation via 
\vspace{-8pt}
\begin{align}
    \mathbf{C}_{a}=\tanh \left(\frac{\mathbf{X}_{a}^T \mathbf{W}_{j a} \mathbf{J}}{\sqrt{d}}\right), \\
    \mathbf{C}_{\mathbf{v}}=\tanh \left(\frac{\mathbf{X}_{v}^T \mathbf{W}_{j v} \mathbf{J}}{\sqrt{d}}\right),   
\end{align}
where $\mathbf{W}_{j a}$ and $\mathbf{W}_{j v}$ are projecting weight matrices for different modalities. The obtained $\mathbf{C}_a$ and $\mathbf{C}_v$ measure the relevance not only across different modalities but also within the same modality. Then, we combine them with their corresponding uni-modal representations and get the fusion features $\mathbf{X}_{j,a}$ and $\mathbf{X}_{j,v}$. Take the audio feature $\mathbf{X}_{j,a}$ as an example, the computational procedure can be summarized as

\vspace{-10pt}
\begin{align}
\mathbf{H}_{a}&=\text{ReLU}\left(\mathbf{W}_{a} \mathbf{X}_{a}+\mathbf{W}_{c a} \mathbf{C}_{a}^T\right), \\
\mathbf{X}_{j, a}&=\mathbf{W}_{h a} \mathbf{H}_{a}+\mathbf{X}_{a},
\end{align}
where $\mathbf{W}_a$, $\mathbf{W}_{c a}$, and $\mathbf{W}_{h a}$ are learnable weight matrices.
$\mathbf{H}_a$ can be denoted as the attention maps of audio modality and $\mathbf{X}_{j,a}$ is the attended features of audio modality.
After computing $X_{j,v}$ via a similar process, we finally concatenate it with $\mathbf{X}_{j,a}$ to form the audio-visual representation for predicting the keyword existence.

\vspace{-10pt}
\section{Experiments}
\label{sec:experiments}
\vspace{-5pt}
\subsection{Database and Evaluation Metrics}
\vspace{-5pt}
We evaluate our proposed system on the audio-visual wake word spotting (WWS) database released in the MISP2021 Challenge. The database contains about 125 hours of audio-visual data recorded in over 30 real-world rooms. The data has three subsets: a training set (47k+ negative samples and 5K+ positive samples), a development set (2k+ negative samples and 600+ positive samples), and an evaluation set (8K+ samples). The audio collection devices include a 6-channel far-field microphone array, a 2-channel mid-field microphone array, and a single-channel near-field microphone, while the video collection devices include a mid-field high-definition camera and a far-field wide-angle camera. 

In this case, if the wake-up word ``Xiao T, Xiao T" is included in the sample, it is considered a positive sample, otherwise, it will be regarded as a negative sample. Following the MISP challenge setup~\cite{misp}, the combination of false reject rate (FRR) and false alarm (FAR) is adopted to evaluate the systems, as follows:
\vspace{-8pt}
\begin{equation}\label{eq:WWS}
\text { Score }=\text{FRR}+\text{FAR}=\frac{N_{\text{FR}}}{N_{\text {wake }}}+\frac{N_{\text{FA}}}{N_{\text {non-wake }}},
\end{equation}
where ${N_{\text{wake}}}$ and ${N_{\text{non-wake}}}$ denote the number of samples including wake-up word or without wake-up word. ${N_{\text{FR}}}$ denotes the number of samples that include the wake word but are not detected by the system and ${N_{\text{FA}}}$ is the number of samples that do not contain the keyword but are detected erroneously. The lower the score, the better the system performance.
\vspace{-10pt}
\subsection{Setups of the Video-Assisted MVDR}
\label{sec:fe}
\vspace{-5pt}
The mask estimate network consists of three 256-dim long short-term memory (LSTM) layers, followed by a full connection (FC) layer and a Sigmoid activation function to estimate IRM. 
Since the view width of the far-field camera is 120 degrees, we equally divide the video frame into six parts horizontally, and each region corresponds to a beam of 20 degrees. The beam index for the input sample is got through the ROI released with the MISP dataset, which is cropped by the internal tools provided by the challenge organizer.

For the training of our mask estimate network, we use the far-field data of the challenge as input and the near-field data as a target. IPD features are calculated among three microphone pairs with indexes (1,4), (2,5), and (3,6). The frame length and hop length used in the short-time Fourier transform (STFT) are set to 32 and 16 ms, respectively. The network is trained for 20 epochs with the Adam optimizer. The initial learning rate is set to 0.001 and will be halved if there is no improvement in the development set.
\vspace{-7pt}
\subsection{Uni-modal Model Pre-training}\label{sec:unimodal}
\vspace{-2pt}

\textbf{Audio processing.} Firstly, we extract 80-dimensional log-Mel Filterbank (FBank) features with 25ms frame length and 10ms frame shift from raw waves. 
Because the frame length of the video is about 4 times larger than the speech, we use a window of 80 frames sliding along the time axis with a stride of 4 to form a sequence of feature blocks. The shape of feature blocks is $(T, 80, 80, 1)$, where $T$ refers to the number of feature blocks and $1$ means the channel dimension.
In addition, speed perturbation, pitch shifting and noise augmentation are also applied as the data augmentation strategies during the training to improve the robustness of the model.

\noindent\textbf{Visual processing.}
We crop the lip regions of the videos according to the ROI provided by MISP organizers and resize them to have a resolution of $112\times112$ with 3 RGB channels. The shape of visual features is $(T, 112, 112, 3)$, where $T$ refers to the number of time steps and $3$ means the channel dimension.
We also use some video-based data augmentation techniques, including frame-wise rotation, flipping horizontally, and color transformation.

\noindent\textbf{Model training.} We employ the same network architecture for both modalities, which consists of five ResNet3D blocks, whose channels are 32, 64, 64, 128, and 256 respectively.
Both models are trained for 15 epochs with Adam optimizer, where the initial learning rate is set to 0.001 and will halve if there is no improvement on the development set. To tackle the imbalance between positive and negative samples, we adopt the weighted CrossEntropy loss (negative:positive=1:5) with the label smoothing method~\cite{labelsmoothing}.
\vspace{-10pt}

\subsection{Joint Fusion Training}
\vspace{-5pt}
During the training of our VE-KWS model, the modality-exclusive parts are initialized by the pre-trained uni-modal models, and the fusion module is randomly initialized. Most of the training setups are the same as Sec~\ref{sec:unimodal}, except the initial learning rate is $1e^{-4}$.
In order to increase the diversity of training data, we randomly replace the video with another video in the same class to simulate more audio-visual input pairs.


\vspace{-10pt}
\subsection{Results and Discussion}
\vspace{-5pt}

Table~\ref{tab:audio} shows ablation studies of the audio-modality model. The baseline represents the model trained without any extra tricks. For the BF method, we estimate the arrival direction through the lip region coordinates and then perform fixed beamforming. Compared with the above methods, our model gives a significant improvement, leading up to 36.50\%/12.02\% relative score reduction over the baseline and 24.12\%/7.59\% over the fixed beamforming system on the Dev/Eval sets, respectively. Furthermore, by incorporating the data augmentation methods into training, we obtain the best performance of 9.78\% on the Eval set but a slight degradation on the Dev set.

\vspace{-5pt}

\begin{table}[!htb]
\centering
\tabcolsep=0.13cm
\caption{Experimental results of the audio-modality models.}
\vspace{-5pt}
\begin{tabular}{clcccccc}
\toprule
 & \multirow{2}{*}{Method} & \multicolumn{3}{c}{Dev( \%)} & \multicolumn{3}{c}{Eval( \%)} \\ \cline{3-8} 
                    &                         & FRR     & FAR     & Score   & FRR     & FAR     & Score    \\ \hline
                  & Baseline                & 6.89    & 8.07    & 14.96   & 8.91    & 5.48    & 14.39    \\
                  & ~~+ BF                & 8.51    & 4.01    & 12.52   & 7.92    & 5.78    & 13.70    \\
                  & ~~+ audio-only MVDR            & 7.52    & 3.97    & 11.49   & 7.61    & 5.52    & 13.13   \\
                  & ~~+ proposed MVDR                & 6.09    & 3.41    & \textbf{9.50}   & 9.42    & 3.22    & 12.64   \\
                  & ~~~~~+Data Aug.         & 6.73    & 3.12    & 9.85    & 5.82    & 3.95    & \textbf{9.78}     \\ \bottomrule
\end{tabular}
\label{tab:audio}
\end{table}

\vspace{-5pt}
Table~\ref{tab:visual} are the ablation studies of the visual-modality model. From the table, we can find that data augmentation can reduce the score significantly. Besides, when using the additional CAS-VSR-W1k dataset~\cite{lrw1000} for pre-training, we can observe further improvement in both sets. However, the results of the visual-modality model have a large gap with the audio-modality system, which verifies our statement mentioned in Sec~\ref{Sec:2-2-2}.

\begin{table}[!htb]
\centering
\tabcolsep=0.13cm
\caption{Experimental results of the visual-modality models.}
\vspace{-5pt}
\begin{tabular}{clcccccc}
\toprule
 & \multirow{2}{*}{Method} & \multicolumn{3}{c}{Dev( \%)} & \multicolumn{3}{c}{Eval( \%)} \\ \cline{3-8} 
                    &                         & FRR     & FAR     & Score   & FRR     & FAR     & Score    \\ \hline
                  & Baseline                & 9.96    & 12.78    & 22.74   & 18.91    & 12.30    & 31.21    \\
                  & ~~+Data Aug.                & 5.45    & 10.14    & 15.59    & 15.94    & 10.43    & 26.37    \\
                  & ~~~~+Pre-train         & 6.73    & 6.68    & \textbf{13.41}    &18.39    & 7.62    &\textbf{26.01}     \\ \bottomrule
\end{tabular}
\label{tab:visual}
\vspace{-10pt}
\end{table}

Table~\ref{tab:av} presents the ablation studies of the audio-visual model. We compare our fusion method with the classic fusion method on the feature level and decision level, respectively. For feature-level fusion, the outputs of uni-modal models are concatenated and fed into fully-connected layers to make the final decision. For decision-level fusion, we simply average the score predicted by each uni-modal model and make the final decision. With the help of the proposed methods, our model shows superior performance over feature- or decision-level fusion models on both sets, achieving up to 44.31\% and 33.10\% relative score reductions, respectively.


\begin{table}[!htb]
\centering
\tabcolsep=0.13cm
\caption{Experimental results of the multi-modal fusion models.}
\vspace{-5pt}
\begin{tabular}{cccccccc}
\toprule
& \multirow{2}{*}{Method} & \multicolumn{3}{c}{Dev( \%)} & \multicolumn{3}{c}{Eval( \%)} \\ \cline{3-8} 
                    &                         & FRR     & FAR     & Score   & FRR     & FAR     & Score    \\ \hline
                  & Decision Fusion               & 1.60    & 4.90    & 6.50   & 2.76    & 5.82    & 8.58   \\
                  & Feature Fusion                & 2.24    & 2.74    & 4.98    & 3.80    & 2.85    & 6.65     \\
                  & VE-KWS (Proposed)          & 1.60    & 2.02    & \textbf{3.62}   & 2.79   & 2.95    & \textbf{5.74}      \\ \bottomrule
\end{tabular}
\label{tab:av}
\vspace{-5pt}
\end{table}

To detail analyze the system performance of different noisy environments, we averagely split the Eval set into three subsets according to the BAK of Dnsmos~\cite{dnsmos}. From the result in Table~\ref{tab:subset}, we can find that the audio-modality system is susceptible to performance degradation when noises exist, resulting in a large gap between noisy and clean sets. However, the visual-modality system is not affected by the noise, performing similarly on different subsets. When fusing audio and visual modalities, our model indeed overcomes the degradation caused by noisy conditions. Table~\ref{tab:other_sys} shows the comparison with other studies in the challenge and our model outperforms the top-ranking systems resulting in a new SOTA performance.


\begin{table}[!htb]
\centering
\caption{Experimental results on the Eval subsets.}
\vspace{-5pt}
\begin{tabular}{cccccc}
\toprule
&\multirow{2}{*}{Method} & \multicolumn{4}{c}{Score( \%)}  \\ \cline{3-6}  
 &                       & All   & Noisy & Medium & Clean \\ \toprule
&Audio                   & 9.78  & 12.29 & 8.77   & 8.31  \\
&Visual                  & 26.01 & 23.78 & 28.41  &26.00  \\
 &VE-KWS (Proposed)                       & 5.74  & 7.18  & 5.61   & 4.47 \\ \bottomrule
\end{tabular}
\label{tab:subset}
\vspace{-15pt}
\end{table}

\begin{table}[!htb]
\centering
\caption{Results comparison with other released systems.}
\label{tab:other_sys}
\vspace{-5pt}
\begin{tabular}{cccc}
\toprule
\multirow{2}{*}{Model} & \multicolumn{1}{l}{\multirow{2}{*}{\begin{tabular}[c]{@{}l@{}}MISP\\ ~Rank~\end{tabular}}} & \multicolumn{2}{c}{Score( \%)} \\ \cline{3-4}
                       & \multicolumn{1}{l}{}                                                                     & Dev           & Test         \\\toprule
CNN-TDNNF+MMI~\cite{misp}          & $1^{th}$                                                                                      & 4.1           & 5.8          \\
ResNet+HMA~\cite{dku}             & $2^{nd}$                                                                                      & 7.27          & 7.1          \\
Transformer+Fusion~\cite{weilai}     & $3^{rd}$                                                                                      & -             & 9.1          \\ \midrule
VE-KWS (Proposed)                 & -                                                                                 & \textbf{3.62}           & \textbf{5.74}        \\ \bottomrule 
\end{tabular}
\vspace{-15pt}
\end{table}

\section{Conclusion}
\label{sec:Conclusion}
\vspace{-5pt}
In this work, we propose a novel visual modality enhanced end-to-end KWS framework (VE-KWS), which utilizes audio-visual information to improve the performance of KWS system. 
Specifically, the model consists of a visual-assisted MVDR that utilizes the lip region information to provide speaker direction information for the beamformer and a joint audio-visual feature fusion model that leverages complementary relationships over multiple modalities. Experimental results on the MISP challenge show that our model achieves a 2.79\% false rejection rate and 2.95\% false alarm rate on the Eval set, outperforming other submitted systems and resulting in a new SOTA performance. Although the MISP challenge does not restrict the real-time factor (RTF) and model size for  submissions, small footprint, and real-time processing are essential for a practical KWS application. Thus, we would like to explore audio-visual KWS systems with low RTF and small footprint in our future work.



\bibliographystyle{IEEE}
\bibliography{strings,refs}

\begin{thebibliography}{10}

\bibitem{kws_0}
Aleksandr Laptev, Roman Korostik, Aleksey Svischev, Andrei Andrusenko, Ivan
  Medennikov, and Sergey Rybin,
\newblock ``You do not need more data: Improving end-to-end speech recognition
  by text-to-speech data augmentation,''
\newblock in {\em Proc. CISP-BMEI}. 2020, pp. 439--444, {IEEE}.

\bibitem{kwsrelate0}
Guoguo Chen, Carolina Parada, and Georg Heigold,
\newblock ``Small-footprint keyword spotting using deep neural networks,''
\newblock in {\em Proc. ICASSP}. 2014, pp. 4087--4091, {IEEE}.

\bibitem{kwsrelate1}
Yiming Wang, Hang Lv, Daniel Povey, Lei Xie, and Sanjeev Khudanpur,
\newblock ``Wake word detection with streaming transformers,''
\newblock in {\em Proc. ICASSP}. 2021, pp. 5864--5868, {IEEE}.

\bibitem{kwsrelate2}
Jingyong Hou, Yangyang Shi, Mari Ostendorf, Mei{-}Yuh Hwang, and Lei Xie,
\newblock ``Mining effective negative training samples for keyword spotting,''
\newblock in {\em Proc. ICASSP}. 2020, pp. 7444--7448, {IEEE}.

\bibitem{hard1}
Yiteng~Arden Huang, Turaj~Zakizadeh Shabestary, and Alexander Gruenstein,
\newblock ``Hotword cleaner: Dual-microphone adaptive noise cancellation with
  deferred filter coefficients for robust keyword spotting,''
\newblock in {\em Proc. ICASSP}. 2019, pp. 6346--6350, {IEEE}.

\bibitem{hard2}
Emre Yilmaz, {\"{O}}zg{\"{u}}r~Bora Gevrek, Jibin Wu, Yuxiang Chen, Xuanbo
  Meng, and Haizhou Li,
\newblock ``Deep convolutional spiking neural networks for keyword spotting,''
\newblock in {\em Proc. Interspeech}. 2020, pp. 2557--2561, {ISCA}.

\bibitem{hard3}
Meng Yu, Xuan Ji, Bo~Wu, Dan Su, and Dong Yu,
\newblock ``End-to-end multi-look keyword spotting,''
\newblock in {\em Proc. Interspeech}. 2020, pp. 66--70, {ISCA}.

\bibitem{convmixer}
Dianwen Ng, Yunqi Chen, Biao Tian, Qiang Fu, and Eng~Siong Chng,
\newblock ``Convmixer: Feature interactive convolution with curriculum learning
  for small footprint and noisy far-field keyword spotting,''
\newblock in {\em Proc. ICASSP}. 2022, pp. 3603--3607, {IEEE}.

\bibitem{emotion}
R.~Gnana Praveen, Wheidima~Carneiro de~Melo, Nasib Ullah, Haseeb Aslam, Osama
  Zeeshan, Th{\'{e}}o Denorme, Marco Pedersoli, Alessandro~L. Koerich, Simon
  Bacon, Patrick Cardinal, and Eric Granger,
\newblock ``A joint cross-attention model for audio-visual fusion in
  dimensional emotion recognition,''
\newblock in {\em Proc. CVPR}. 2022, pp. 2485--2494, {IEEE}.

\bibitem{scene}
Han Lei and Ning Chen,
\newblock ``Audio-visual scene classification based on multi-modal graph
  fusion,''
\newblock in {\em Proc. Interspeech}. 2022, pp. 4157--4161, {ISCA}.

\bibitem{avkws}
Liliane Momeni, Triantafyllos Afouras, Themos Stafylakis, Samuel Albanie, and
  Andrew Zisserman,
\newblock ``Seeing wake words: Audio-visual keyword spotting,''
\newblock in {\em Proc. BMVC}, 2020.

\bibitem{misp}
Hang Chen, Hengshun Zhou, Jun Du, Chin{-}Hui Lee, Jingdong Chen, Shinji
  Watanabe, Sabato~Marco Siniscalchi, Odette Scharenborg, Diyuan Liu, Bao{-}Cai
  Yin, Jia Pan, Jianqing Gao, and Cong Liu,
\newblock ``The first multimodal information based speech processing (misp)
  challenge: Data, tasks, baselines and results,''
\newblock in {\em Proc. ICASSP}. 2022, pp. 9266--9270, {IEEE}.

\bibitem{dku}
Ming Cheng, Haoxu Wang, Yechen Wang, and Ming Li,
\newblock ``The {DKU} audio-visual wake word spotting system for the 2021
  {MISP} challenge,''
\newblock in {\em Proc. ICASSP}. 2022, pp. 9256--9260, {IEEE}.

\bibitem{weilai}
Yanguang Xu, Jianwei Sun, Yang Han, Shuaijiang Zhao, Chaoyang Mei, Tingwei Guo,
  Shuran Zhou, Chuandong Xie, Wei Zou, and Xiangang Li,
\newblock ``Audio-visual wake word spotting system for {MISP} challenge 2021,''
\newblock in {\em Proc. ICASSP}. 2022, pp. 9246--9250, {IEEE}.

\bibitem{doi1}
Zhong{-}Qiu Wang and DeLiang Wang,
\newblock ``Combining spectral and spatial features for deep learning based
  blind speaker separation,''
\newblock {\em {IEEE} {ACM} Trans. Audio Speech Lang. Process.}, vol. 27, no.
  2, pp. 457--468, 2019.

\bibitem{doi2}
Rongzhi Gu, Lianwu Chen, Shi{-}Xiong Zhang, Jimeng Zheng, Yong Xu, Meng Yu, Dan
  Su, Yuexian Zou, and Dong Yu,
\newblock ``Neural spatial filter: Target speaker speech separation assisted
  with directional information,''
\newblock in {\em Proc. Interspeech}. 2019, pp. 4290--4294, {ISCA}.

\bibitem{doi3}
Fahimeh Bahmaninezhad, Jian Wu, Rongzhi Gu, Shi{-}Xiong Zhang, Yong Xu, Meng
  Yu, and Dong Yu,
\newblock ``A comprehensive study of speech separation: Spectrogram vs waveform
  separation,''
\newblock in {\em Proc. Interspeech}. 2019, pp. 4574--4578, {ISCA}.

\bibitem{mvdr}
Takuya Yoshioka, Yan Huang, Aviv Hurvitz, Li~Jiang, Sharon Koubi, Eyal Krupka,
  Ido Leichter, Changliang Liu, Partha Parthasarathy, Alon Vinnikov, Lingfeng
  Wu, Igor Abramovski, Xiong Xiao, Wayne Xiong, Huaming Wang, Zhenghao Wang,
  Jun Zhang, Yong Zhao, Tianyan Zhou, Cem Aksoylar, Zhuo Chen, Moshe David,
  Dimitrios Dimitriadis, Yifan Gong, Ilya Gurvich, and Xuedong Huang,
\newblock ``Advances in online audio-visual meeting transcription,''
\newblock in {\em Proc. ASRU}. 2019, pp. 276--283, {IEEE}.

\bibitem{resnet3d}
Zhaofan Qiu, Ting Yao, and Tao Mei,
\newblock ``Learning spatio-temporal representation with pseudo-3d residual
  networks,''
\newblock in {\em Proc. ICCV}. 2017, pp. 5534--5542, {IEEE} Computer Society.

\bibitem{resnet3dar}
Kensho Hara, Hirokatsu Kataoka, and Yutaka Satoh,
\newblock ``Learning spatio-temporal features with 3d residual networks for
  action recognition,''
\newblock in {\em Proc. ICCV}. 2017, pp. 3154--3160, {IEEE} Computer Society.

\bibitem{speaker}
Danwei Cai, Xiaoyi Qin, and Ming Li,
\newblock ``Multi-channel training for end-to-end speaker recognition under
  reverberant and noisy environment,''
\newblock in {\em Proc. Interspeech}. 2019, pp. 4365--4369, {ISCA}.

\bibitem{attisneed}
Ashish Vaswani, Noam Shazeer, Niki Parmar, Jakob Uszkoreit, Llion Jones,
  Aidan~N. Gomez, Lukasz Kaiser, and Illia Polosukhin,
\newblock ``Attention is all you need,''
\newblock in {\em Proc. NIPS}, 2017, pp. 5998--6008.

\bibitem{bilinear}
Joshua~B Tenenbaum and William~T Freeman,
\newblock ``Separating style and content with bilinear models,''
\newblock {\em Neural computation}, vol. 12, no. 6, pp. 1247--1283, 2000.

\bibitem{labelsmoothing}
Rafael M{\"{u}}ller, Simon Kornblith, and Geoffrey~E. Hinton,
\newblock ``When does label smoothing help?,''
\newblock in {\em Proc. NIPS}, 2019, pp. 4696--4705.

\bibitem{lrw1000}
Shuang Yang, Yuanhang Zhang, Dalu Feng, Mingmin Yang, Chenhao Wang, Jingyun
  Xiao, Keyu Long, Shiguang Shan, and Xilin Chen,
\newblock ``{LRW-1000:} {A} naturally-distributed large-scale benchmark for lip
  reading in the wild,''
\newblock in {\em Proc. FG}. 2019, pp. 1--8, {IEEE}.

\bibitem{dnsmos}
Chandan~KA Reddy, Vishak Gopal, and Ross Cutler,
\newblock ``Dnsmos: A non-intrusive perceptual objective speech quality metric
  to evaluate noise suppressors,''
\newblock in {\em Proc. ICASSP}. 2021, pp. 6493--6497, IEEE.

\end{thebibliography}

\end{document}